\documentclass[intlimits,twoside,a4paper]{article}

\usepackage{graphicx}

\usepackage[cp1251]{inputenc}

\usepackage[eqsecnum]{cmpj3}

\issue{2017}{20}{2}{23401}
\doinumber{10.5488/CMP.20.23401}

\title[Enhancement of local field in core-shell orientation of ellipsoidal nanoinclusions]%
{Enhancement of local electric field in core-shell orientation of ellipsoidal metal/dielectric nanoparticles%
\thanks{E-mail: abdulhayi.abdella@aau.edu.et}}
\author[A.A. Ismail, A.V. Gholap, Y.A. Abbo]{A.A. Ismail\refaddr{label1}, A.V. Gholap\refaddr{label1}, Y.A. Abbo\refaddr{label2}}
\addresses{
\addr{label1} Department~of~Physics,~Addis~Ababa~University,~P~O~Box~1176,~Addis~Ababa,~Ethiopia
\addr{label2} Department~of~Physics,~Wollega~University,~P~O~Box~395,~Nekemt,~Ethiopia}

\date{Received October 20, 2016, in final form January 10, 2017}

\begin{document}

\maketitle

\begin{abstract}
In this paper it is shown that the enhancement factor of the local
electric field in metal covered ellipsoidal nanoparticles embedded
in a dielectric host matrix has two maxima at two different frequencies.
The second maximum for the metal covered inclusions with large
dielectric core (small metal fraction $p$) is comparatively large.
This maximum strongly depends on the depolarization factor of the core
$L_{z}^{(1)}$, keeping that of the shell $L_{z}^{(2)}$ constant and
is less than $L_{z}^{(1)}$. If the frequency of the external
radiation approaches the frequency of surface plasmons of a metal,
the local field in the particle considerably increases. The importance of maximum value of enhancement factor $|A|^{2}$ of the ellipsoidal inclusion is emphasized in the case where the dielectric core exceeds metal fraction of the inclusion. The results
of numerical computations for typical small silver particles are
presented graphically.
\pacs  42.65.Pc, 42.79.Ta, 78.67.Sc, 78.67.-n
\end{abstract}

\section{Introduction}

The enhancement of the local electric field of the incident
electromagnetic radiation in the composites of metal covered
nanoparticles with dielectric core is of great  importance due to different possible applications such as surface enhanced Raman spectroscopy \cite{Ref1}, metal enhanced  florescence \cite{Ref2},
quantum electrodynamics \cite{Ref4,Ref5}, nonlinear optical effect \cite{Ref6}, quantum optomechanics \cite{Ref7}, optical sensors \cite{Ref8} and nano-optical
tweezers \cite{Ref9}. It is known
that the local electric field in the inclusions can be considerably
enhanced if a frequency of the incident radiation is close to the
surface plasmon frequency \cite{Nee89}. This problem was studied in connection
with the optically induced bistability \cite{Ref10,Ref11} and it is
accepted that such an enhancement takes place only on
one resonant frequency. It is clear that the nonlinear part of
the dielectric function (DF) is important only if the electric
fields are comparable with the inner atomic fields. At present, such
fields may be achieved by laser radiation. Another interesting
property of a pure metal and metal-covered dielectric small
particles is an abnormal enhancement of the local field, when the
frequency of the incident electromagnetic wave approaches the
surface plasmon frequency of the metal \cite{Bur11}. The fact that the surface plasmon (SP) strongly depends on size, shape, distribution
of metal nanoparticles as well as on the surrounding dielectric matrix offers an
opportunity for manufacturing new promising nonlinear materials, nanodevices
and optical elements. The ellipsoidal shape does represent the most general
geometry suitable for many practical applications: in
particular, it allows us to analyze two important limiting
cases, namely the spherical and the cylindrical ones. The existence of a two-peak value structure of the frequency dependence of the enhancement factor was first presented by Sisay and Mal’nev \cite{Sis12} for a composite with metal coated spherical nanoinclusions.
Under this context, the present investigation provides a
very general conceptual framework, including those specific
cases previously investigated. A detailed theoretical and
numerical analysis of the local field enhancement in small metal
covered ellipsoidal inclusions in the electrostatic approximation is the aim of
this study. In section~\ref{sec-2} and~\ref{sec-3}, we analyze the distribution of electric potential in a coated ellipsoidal metal nanoparticle when the incident electric field is parallel to
one of the ellipsoid axes  ($z$-axis), and the enhancement factor of local field
inside a metal covered ellipsoidal dielectric core embedded into a
dielectric matrix, respectively. Lastly, in section~\ref{sec-4}, the results of numerical
calculations are illustrated graphically. In the conclusion, we
summarize the main results of the paper.

\section{Electric potentials distribution in a coated ellipsoidal metal nanoparticle}
\label{sec-2}
\begin{figure}[!b]
	\centerline{\includegraphics[width=1\textwidth]{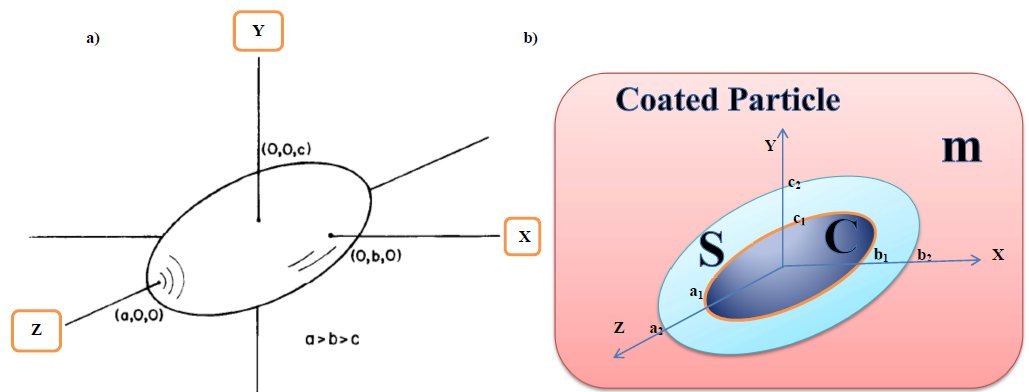}}
	\caption{ (Color online) a) Ellipsoidal particle.  b) Metal coated  ellipsoidal particle with internal
		dielectric core (c), external shell (s) embedded into a given matrix
		(m).} \label{fig-smp1}
\end{figure}

The most general smooth particle (the one without edges or corners) of regular shape of an ellipsoidal coordinates \cite{url1}, with semiaxes $a > b > c$ [figure~\ref{fig-smp1}~(a)], can be obtained by considering the surface which is specified by
\begin{eqnarray}
\dfrac{x^{2}}{a^{2}}+\frac{y^{2}}{b^{2}}+\dfrac{z^{2}}{c^{2}}=1
\end{eqnarray}
without loss of generality, considering a family of curves defined by
\[f(q)\equiv\dfrac{x^{2}}{a^{2}+q}+\dfrac{y^{2}}{b^{2}+q}+\dfrac{z^{2}}{c^{2}+q}-1=0\]
for $q>-c^{2}$, $f(q)=0$ defines an ellipsoid.

Consider a confocal core-shell ellipsoid shown in
figure~\ref{fig-smp1}~(b), which can represent a wide range of shapes from
disks to rods.  The principal semiaxes are $a_{1}$, $b_{1}$, and $c_{1}$ for the
core surface and $a_{2}$, $b_{2}$, and $c_{2}$ for the outer shell surface.
Any confocal ellipsoidal surface can be expressed by
\begin{eqnarray}
\label{eqn-2.2}
 \dfrac{x^{2}}{a^{2}_{1}+q}+\dfrac{y^{2}}{b^{2}_{1}+q}+\dfrac{z^{2}}{c^{2}_{1}+q}=1,    \qquad     (a_{1}>b_{1}>c_{1}).
\end{eqnarray}
  This equation, a cubic in $q$, has three real roots $\xi$, $\eta$, and $\zeta$ that
  define the ellipsoidal coordinates.  The coordinate $ \xi$ is normal
  to the surface.
  The variables $\eta$ and  $\zeta$  are the parameters of confocal hyperboloids and as such serve to measure the position on any ellipsoid $\xi=\text{constant}$. In other words, each ellipsoidal surface is
  defined by a constant $ \xi$.  Therefore, $\xi=0$ is the equation of the surface of inner ellipsoid and $\xi=t$ is that of the surface of the outer ellipsoid, where $a_{1}^{2}+t=a_{2}^{2}$,  $b_{1}^{2}+t=b_{2}^{2}$,  $c_{1}^{2}+t=c_{2}^{2}$.  For a given $(x,y,z)$,  if we assume $x>0$, $y>0$, $z>0$, there is a one to one correspondence between $(x,y,z)$ and the three largest roots $(\xi,\eta,\zeta)$. This implies that the transformation to rectangular coordinates is obtained by solving
  equation~(\ref{eqn-2.2}) simultaneously for $x$, $y$, $z$,  which shows that an expression for the $z$ will be given~as:
   \begin{eqnarray}
   \label{eqn-2.3}
   z=\left[\frac{(c_{1}^{2}+\xi)(c_{1}^{2}+\eta)(c_{1}^{2}+\zeta)}{(a_{1}^{2}-c_{1}^{2})(b_{1}^{2}-c_{1}^{2})}\right]^{1/2}.
   \end{eqnarray}
   We assume that the uniform electrostatic field $\textbf{E}_{0}$ is directed along the $z$-axis. Then, the external field potential can be written in the form
\begin{eqnarray}
\label{eqn-2.4}
\Phi_{0}=-E_{0}z=-E_{0}\left[\frac{(c_{1}^{2}+\xi)(c_{1}^{2}+\eta)(c_{1}^{2}+\zeta)}{(a_{1}^{2}-c_{1}^{2})(b_{1}^{2}-c_{1}^{2})}\right]^{1/2}
=-E_{0}F_{1}(\xi)G(\eta,\zeta),
\end{eqnarray}
where we substitute $F_{1}(\xi)=(c_{1}^{2}+\xi)^{1/2}$ and $G(\eta,\zeta)=\{(c_{1}^{2}+\eta)(c_{1}^{2}+\zeta)/[(a_{1}^{2}-c_{1}^{2})(b_{1}^{2}-c_{1}^{2})]\}^{1/2}$ from equation~(\ref{eqn-2.3}).

Let us assume an electrostatic approximation in which the wavelength of the incident electromagnetic wave
is much greater than the typical size of the inclusion. The
distribution of electric potentials between the interfaces of an ellipsoidal metal coated nanoparticle can be expressed as: (i)~$\Phi_{\text c}$ in the dielectric core, (ii)~$\Phi_{\text s}$ in the metal cover shell and (iii)~$\Phi_{\text m}$ in the embedded dielectric matrix.   Under the action of a
 constant external electric field $\textbf{E}_{0}$,  they can be described by the following expressions \cite{Boh83}
\begin{align}
\label{eqn-2.5}
\nonumber
&\Phi_{\text c}=K_{1}F_{1}(\xi)G(\eta,\zeta), \qquad -c_{1}^{2}<\xi<0,\\
\nonumber
&\Phi_{\text s}=[K_{2}F_{1}(\xi)+K_{3}F_{2}(\xi)]G(\eta,\zeta), \qquad 0\leqslant\xi<t, \\
&\Phi_{\text m}=\Phi_{0}+\Phi_{\text p}\,, \qquad t\leqslant\xi<\infty,
\end{align}
which are the solutions of the Laplace's equation in ellipsoidal coordinates stated as:
\begin{align}
\label{eqn-2.6}
\nabla^{2}\Phi_{i}&=(\eta-\zeta)f(\xi)\dfrac{\partial}{\partial\xi}\left[f(\xi)\dfrac{\partial\Phi_{i}}{\partial\xi}\right]
+(\zeta-\xi)f(\eta)\dfrac{\partial}{\partial\eta}\left[f(\eta)\dfrac{\partial\Phi_{i}}{\partial\eta}\right] \nonumber\\
&\quad+(\xi-\eta)f(\zeta)\dfrac{\partial}{\partial\zeta}\left[f(\zeta)\dfrac{\partial\Phi_{i}}{\partial\zeta}\right]=0,
\end{align}
 where $f(q)=[(a^{2}+q)(b^{2}+q)(c^{2}+q)]^{1/2}$, here $q$ stands for $\xi$, $\eta$, $\zeta$ in the function of $ f(\xi)$, $f(\eta)$, $f(\zeta)$ of the ellipsoidal coordinates in equation~(\ref{eqn-2.6}) above. The subscript ``$i$'' indicates the interface potentials in the dielectric core $i=\text{c}$, metal shell $i=\text{s}$, and in the host matrix $i=\text{m}$, respectively.  The potential~$\Phi_{\text m}$ in the surrounding medium of equation~(\ref{eqn-2.5}) is the sum of $\Phi_{0}$ and the perturbing
potential $\Phi_{\text p}$ of the particle which is given by
\begin{equation}
\label{eqn-2.7}
\Phi_{\text p}=K_{4}F_{2}(\xi)G(\eta,\zeta),
\end{equation}
where $F_{2}(\xi)=F_{1}(\xi)\int_{\xi}^{\infty}\rd q/(c_{1}^{2}+q)f_{1}(q)$, with the property    $ \lim_{\xi\rightarrow\infty}F_{2}(\xi)=0$; $f_{1}(q)=[(a_{1}^{2}+q)(b_{1}^{2}+q)(c_{1}^{2}+q)]^{1/2}$, and $ K_{1}$, $K_{2}$, $K_{3}$, $K_{4}$ are unknown
constants, to be determined by the boundary conditions specified below.
Therefore, the potential $\Phi_{\text m}$, in the surrounding medium can be expressed by putting equations~(\ref{eqn-2.4})  and (\ref{eqn-2.7}) in (\ref{eqn-2.5}) as:
\begin{eqnarray}
 \nonumber
\Phi_{\text m}=[-E_{0}F_{1}(\xi)+K_{4}F_{2}(\xi)]G(\eta,\zeta),   \qquad   t\leqslant\xi<\infty.
 \end{eqnarray}
The boundary conditions for the potentials can be found from the continuity conditions of the potentials themselves:
\begin{align}
\label{eqn-2.8}
\nonumber
\Phi_{\text c}&=\Phi_{\text s} \qquad\text{at} \qquad \xi=0, \\
\Phi_{\text s}&=\Phi_{\text m} \qquad\text{ at}  \qquad \xi=t,
\end{align}
and the normal components of the electric
displacement vector:
\begin{align}
\nonumber
\varepsilon_{1}\frac{\partial\Phi_{\text c}}{\partial\xi}&=\varepsilon_{2}\frac{\partial\Phi_{\text s}}{\partial\xi}
\qquad \text{ at}  \qquad \xi=0,\\
\label{eqn-2.9}
\varepsilon_{2}\frac{\partial\Phi_{\text s}}{\partial\xi}&=\varepsilon_{\text m}\frac{\partial\Phi_{\text m}}{\partial\xi}  \qquad\text{ at}  \qquad\xi=t.
\end{align}
The unknown coefficients $K_{1}$, $K_{2}$, $K_{3}$, $K_{4}$ can be determined by substituting expressions of equation~(\ref{eqn-2.5})  in the system of equations~(\ref{eqn-2.8})  and (\ref{eqn-2.9})  at the boundaries of  dielectric core-metal  and metal-host matrix interfaces, and solving  simultaneously, we obtain a system of linear algebraic equations as listed below:
\begin{align}
K_{1}&=-\frac{\varepsilon_{\text m}\varepsilon_{2}}{Q}E_{0}\,,\\
K_{2}&=-\frac{\varepsilon_{\text m}\big[(\varepsilon_{1}-\varepsilon_{2})L_{z}^{(1)}+\varepsilon_{2}\big]}{Q}E_{0}\,,\\
K_{3}&=\frac{a_{1}b_{1}c_{1}\varepsilon_{\text m}(\varepsilon_{1}-\varepsilon_{2})}{2Q}E_{0}\,,\\
K_{4}&=-\frac{a_{2}b_{2}c_{2}}{2Q}\left\{f(\varepsilon_{1}-\varepsilon_{2})\left[L_{z}^{(2)}(\varepsilon_{2}-\varepsilon_{\text m})-\varepsilon_{2}\right]-(\varepsilon_{2}-\varepsilon_{\text m})\left[L_{z}^{(1)}(\varepsilon_{1}-\varepsilon_{2})+\varepsilon_{2}\right]\right\}E_{0}\,,
\end{align}
where  $Q=p\Delta$,  and
\begin{eqnarray}
\Delta=\varepsilon_{2}^{2}\left[1+\frac{L_{z}^{(2)}-L_{z}^{(1)}}{p}\right]-q\varepsilon_{2}+\varepsilon_{1}\varepsilon_{\text m}\,.
\end{eqnarray}
Here,
 $q=\varepsilon_{1}[1-L_{z}^{(1)}/p]+\varepsilon_{\text m}\{[L_{z}^{(2)}-1]/p+1\}$, and
$ p=fL_{z}^{(2)}[L_{z}^{(2)}-1]-L_{z}^{(1)}[L_{z}^{(2)}-1]$
is a metal fraction in the inclusion which is expressed by the volume fraction
$f=a_{1}b_{1}c_{1}/(a_{2}b_{2}c_{2})$ of the core
into the whole inclusion, that is,  the fraction of the total particle of the volume occupied by the inner ellipsoid. The variables  $ L_{z}^{(1)}$ and
$L_{z}^{(2)}$ are the geometrical factors for the inner and outer
confocal ellipsoids, and, $\varepsilon_{1}$, $\varepsilon_{2}$, and
$\varepsilon_{\text m}$ are the dielectric functions (DFs) of the core,
metal shell, and the host matrix (the surrounding medium), respectively.
Since  we assume that a uniform, parallel electric field $\textbf{E}_{0}$ is directed
along the $z$-axis, and is thus along the major semi-axis of the
ellipsoid ($a$), the local field $E_{\text{loc}}$ in the dielectric core of the
inclusion can be obtained with the help of the relation,
$E_{\text{loc}}=-\nabla\Phi$.   The corresponding local fields in each
portion between the interfaces are; $E^{\text c}_{\text{loc}}$ of the core, $E^{\text s}_{\text{loc}}$ of the shell and $E^{\text m}_{\text{loc}}$ of the surrounding medium of metal coated inclusion. They can be presented with the relation
\begin{eqnarray}
&&
\label{eqn-2.16}
E^{\text c}_{\text{loc}}=-\nabla\Phi_{\text c}=AE_{0}\,,\\&&
E^{\text s}_{\text{loc}}=-\nabla\Phi_{\text s}=BE_{0}+CE_{0}\,,\\&&
E^{\text m}_{\text{loc}}=-\nabla\Phi_{\text m}=E_{0}+DE_{0} \, ,
\end{eqnarray}
the factors that relate the local fields with the external incident
electric field between the interfaces are given as below:
\begin{align}
\label{eqn-2.19}
A&=\frac{\varepsilon_{2}\varepsilon_{\text m}}{Q}\,,\\
B&=\frac{\varepsilon_{\text m}\big[(\varepsilon_{1}-\varepsilon_{2})L_{z}^{(1)}+\varepsilon_{2}\big]}{Q}\,,\\
C&=-\frac{a_{1}b_{1}c_{1}\varepsilon_{\text m}(\varepsilon_{1}-\varepsilon_{2})}{2Q}\int_{\xi}^{\infty}\frac{\rd q}{(c_{1}^{2}+q)f_{1}(q)}\,,\\
\label{eqn-2.22}
D&=\frac{a_{2}b_{2}c_{2}}{2Q}\left[(\varepsilon_{2}-\varepsilon_{\text m})\left\{\varepsilon_{2}+(\varepsilon_{1}-\varepsilon_{2})\left[L_{z}^{(1)}-fL_{z}^{(2)}\right]\right\}+f\varepsilon_{2}(\varepsilon_{1}-\varepsilon_{2})\right]\int_{\xi}^{\infty}
\frac{\rd q}{(c_{1}^{2}+q)f_{1}(q)}\,.
\end{align}
It is important to remark that the external field is completely
controlled by the coefficient $K_{4}$ (or/and $D$). At sufficiently large distances from the particle, the perturbing potential in equation~(\ref{eqn-2.7})  is negligible i.e., when  $\xi\gg a_{2}^{2}$, therefore, we require that  $\lim_{\xi\rightarrow\infty}\Phi_{\text p}=0$. We note that at distances $r$ from the origin which are much greater than the largest semi-axis of the shell $a_{2}$ to any point on the ellipsoid $\xi$,  then $x^{2}+y^{2}+z^{2}=\xi\simeq r^{2}$; the integral in equation~(\ref{eqn-2.7}), that can enter the constant $D$ of expression in equation~(\ref{eqn-2.22})  is approximately:
\begin{eqnarray}
\int_{\xi}^{\infty}\frac{\rd\xi}{(c_{1}^{2}+\xi)f_{1}(\xi)} \simeq \int_{\xi}^{\infty} \dfrac{\rd\xi}{\xi^{5/2}}=\dfrac{2}{3}\xi^{-3/2},  \qquad \left(\xi \simeq r^{2}\gg a_{2}^{2}\right),
\end{eqnarray}
and, therefore, the potential $ \Phi_{\text p}$ is given
\begin{eqnarray}
\label{eqn-2.24}
\Phi_{\text p}\sim \dfrac{E_{0}\cos \theta}{r^{2}}\frac{a_{2}b_{2}c_{2}}{3Q}\left[(\varepsilon_{2}-\varepsilon_{\text m})\left\{\varepsilon_{2}+(\varepsilon_{1}-\varepsilon_{2})\left[L_{z}^{(1)}-fL_{z}^{(2)}\right]\right\}+f(\varepsilon_{1}-\varepsilon_{2})\varepsilon_{2}\right],  \quad (r\gg a_{2}),
\end{eqnarray}
since the potential of ideal dipole is given by $\Phi=P\cos\theta/(4\piup\varepsilon_{\text m}r^{2})$,  we can recognize the equation~(\ref{eqn-2.24})  as the potential of a dipole with the moment
\begin{eqnarray}
P=4\piup\varepsilon_{\text m}\frac{a_{2}b_{2}c_{2}}{3Q}\left[(\varepsilon_{2}-\varepsilon_{\text m})\left\{\varepsilon_{2}+(\varepsilon_{1}-\varepsilon_{2})\left[L_{z}^{(1)}-fL_{z}^{(2)}\right]\right\}+f(\varepsilon_{1}-\varepsilon_{2})\varepsilon_{2}\right].
\end{eqnarray}
Therefore, this yields the polarizability
\begin{eqnarray}
\label{eqn-2.26}
\alpha_{z}=\dfrac{\upsilon\left[(\varepsilon_{2}-\varepsilon_{\text m})\left\{\varepsilon_{2}+(\varepsilon_{1}-\varepsilon_{2})\left[L_{z}^{(1)}-fL_{z}^{(2)}\right]\right\}+f(\varepsilon_{1}-\varepsilon_{2})\varepsilon_{2}\right]}{Q}\,,
\end{eqnarray}
where $\upsilon=4\piup a_{2}b_{2}c_{2}/3$ is the volume of the particle, and $f=a_{1}b_{1}c_{1}/(a_{2}b_{2}c_{2})$ is the volume fraction. Here, it may be mentioned that, letting $q=\xi+t$ to solve the integral that will be substituted by the geometrical factors of the depolarization which is given by
\begin{eqnarray}
L^{k}_{z}=\frac{a_{k}b_{k}c_{k}}{2}\int_{0}^{\infty}\frac{\rd q}{(c_{k}^{2}+q)f_{k}(q)}\,,  \qquad (k=1,2),
\end{eqnarray}
the expression in equation~(\ref{eqn-2.26})  is equivalent to equation~(\ref{eqn-2.22}).
Therefore, equation~(\ref{eqn-2.22}) coincides with the corresponding result shown in \cite{Boh83} for the polarizability of a coated ellipsoid.
The coefficients $A$, $B$, $C$, $D$ are consistent with the \emph{coated sphere} that can be verified by putting $L_{z}^{(1)}=L_{z}^{(2)}=1/3 $, $Q=2p\Delta/9$, as shown by Sisay and Mal’nev \cite{Sis12}.

\section{Resonant frequencies and enhancement factor of local field in metal covered ellipsoidal inclusion}
\label{sec-3}
Among equations~(\ref{eqn-2.19})--(\ref{eqn-2.22}), we need only the coefficients $A$ and $D$ that enter the
potential of the local field in the inclusion ``\emph{core}'' and the induced
dipole moment of the inclusion. Let us consider that the dielectric function of metal $\varepsilon_{2} $ is chosen to be in Drude form \cite{Abb16},
\begin{equation}
\varepsilon_{2}=\varepsilon_{\infty}-\frac{1}{z(z+\ri\gamma)}\,.
\end{equation}
Here, we introduced dimensionless frequencies $z=\frac{\omega}{\omega_{\text p}}$, and $ \gamma=\frac{\nu}{\omega_{\text p}}$
($\omega$ and $\omega_{\text p}$ are the frequency of the incident
radiation and the plasma frequency of the metal shell,
respectively; $\nu$ is the electron collision frequency). $\varepsilon_{\infty}$ is a constant that can be a function of the
frequency and depends on the type of a metal.
The real $\varepsilon_{2}' $ and imaginary $\varepsilon_{2}'' $
parts of dielectric function are given by
\begin{eqnarray}
\varepsilon_{2}'=\varepsilon_{\infty}'-\frac{1}{z^{2}+\gamma^{2}}
\qquad \text{and} \qquad
{\varepsilon_{2}''=\varepsilon_{\infty}''+\frac{\gamma}{z(z^{2}+\gamma^{2})}}\,.
\end{eqnarray}
The dielectric function of the inclusion core $ \varepsilon_{1} $, in general case, includes a non-linear part
with respect to the local field.
\begin{eqnarray}
\label{eqn-3.3}
\varepsilon_{1}=\varepsilon_{10}+\chi|\textbf{E}|^{2},
\end{eqnarray}	
where $ \varepsilon_{10}$ is the linear part of DF, $\chi$ is the nonlinear Kerr
coefficient,  $|\textbf{E}|$
is the local field in the core. For week incident fields $|\textbf{E}|\ll \varepsilon_{10}$, the local field is presented as in equation~(\ref{eqn-2.16})   $ E^{\text c}_{\text{loc}}=AE_{0}$. We call  $A$ the enhancement factor, which is in general a \emph{complex} function. It would be convenient to deal with the
real quantity $|A|^{2}$, which can be presented as follows
\begin{eqnarray}
\label{eqn-3.4}
|A|^{2}=\frac{\varepsilon_{\text m}^{2}}{p^{2}}\frac{\varepsilon_{2}'^{2}+\varepsilon_{2}''^{2}}{\left\{\left(\varepsilon_{2}'^{2}-\varepsilon_{2}''^{2}\right)\left[1+\frac{L_{z}^{(2)}-L_{z}^{(1)}}{p}\right]
	-q\varepsilon'_{2}+\varepsilon_{1}\varepsilon_{\text m}\right\}^{2}+\varepsilon_{2}''^{2}\left\{2\varepsilon_{2}'\left[1+\frac{L_{z}^{(2)}-L_{z}^{(1)}}{p}\right]-q\right\}^{2}}\,.
\end{eqnarray}
For the sake of simplicity, we ignore the imaginary parts of
$\varepsilon_{1}$ and $\varepsilon_{\text m}$.

For an analytic analysis, let us consider an ideal case when a decay of the plasma
vibrations is extremely small
$\gamma\ll 1$.
In this case, the second term in the denominator of equation~(\ref{eqn-3.4})  is proportional to
$\varepsilon_{2}''^{2}\sim \gamma^{2}$ which is very small. Therefore, maximum of the enhancement factor $ |A|^{2}$ corresponds to zero of the first term in the denominator of~(\ref{eqn-3.4}). This condition gives the quadratic equation
with respect to $\varepsilon_{2}'$,
\begin{equation}
\varepsilon_{2}'^{2}\left[1+\frac{L_{z}^{(2)}-L_{z}^{(1)}}{p}\right]
-q\varepsilon'_{2}+\varepsilon_{1}\varepsilon_{\text m}=0.
\end{equation}
It has two roots, and we
obtain two different resonant frequencies $z_{\text r}$.
In the real inclusions, $ \gamma$
is not extremely small but finite.
The behavior of $|A|^{2}$ as a function $z$ in this case
can be analyzed only numerically. The results of this study are presented in the following section.

\section{Numerical results and discussion}
\label{sec-4}
\begin{figure}[!t]
	\centerline{\includegraphics[width=0.65\textwidth]{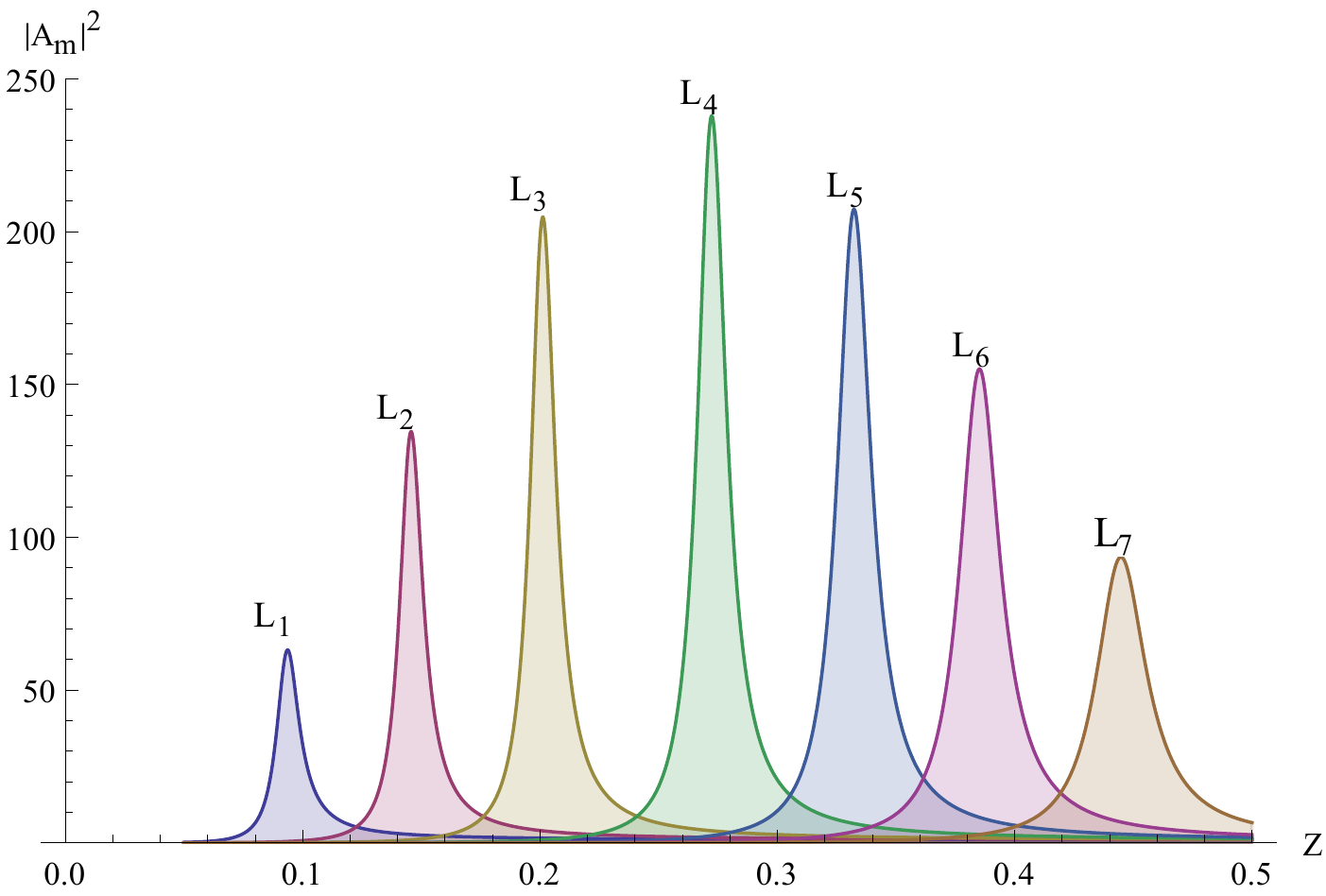}}
\vspace{-2mm}
	\caption{(Color online) The enhancement factor $|A_{\text m}|^{2}$ for a small silver particle
		versus a function of the dimensionless frequency $z$ at different $L$
		($L_{1} = 0.02$; $L_{2} = 0.05$; $L_{3} = 0.1$, $L_{4} = 0.2$, $L_{5} =
		0.33$, $L_{6} = 0.5$, $L_{7} = 0.8)$ with parameters of the particle
		$\varepsilon_{\text m} = 2.25$, $\varepsilon_{\infty}' = 4.5$,
		$\varepsilon_{\infty}'' = 0.16$, $\omega_{\text p} = 1.46\cdot10^{16}$,
		$\nu = 1.68\cdot10^{14}$, $\gamma = 1.15 \cdot10^{-2} $.} \label{fig-smp2}
\end{figure}

We start our numerical calculations with the enhancement factor of a \emph{pure} metal particle
$|A_{\text m}|^{2}$.
The numerical values of the dielectric functions of the composite used in this section are taken from \cite{Bur11,Sis12,Abb16}.
$|A_{\text m}|^{2}$ can be obtained from equation~(\ref{eqn-2.19})  by
setting $p = 1$ and making substitution $\varepsilon_{1} \rightarrow
\varepsilon_{2}$, and $L_{z}^{(1)}=L_{z}^{(2)}=L$ as well
\begin{equation}
\label{eqn-4.1}
|A_{\text m}|^{2}=\frac{\varepsilon_{\text m}^{2}}{[L(\varepsilon_{2}'-\varepsilon_{\text m})+\varepsilon_{\text m}]^{2}+L^{2}\varepsilon_{2}''^{2}}\,,
\end{equation}
where the subscript ``m'' indicates pure metal inclusion. In figure~\ref{fig-smp2}, we present this quantity versus $z$, and one can see that the enhancement factor $|A_{\text m}|^{2}$ in the physically interesting range of
parameters sharply depends on the frequency of an incident
electromagnetic wave $\omega$ and weakly depends on the
depolarization factor decreasing with $L$. It can easily be seen
from equation~(\ref{eqn-4.1})  that $|A_{\text m}|^{2}$ = 1 as $L \rightarrow 0$. The maximum
value of $|A_{\text m}|^{2}$ at the constant values used for numerical calculations of the particle
and the host matrix \cite{Bur11} is around $250$.

\begin{figure}[!t]
	\centerline{\includegraphics[width=0.65\textwidth]{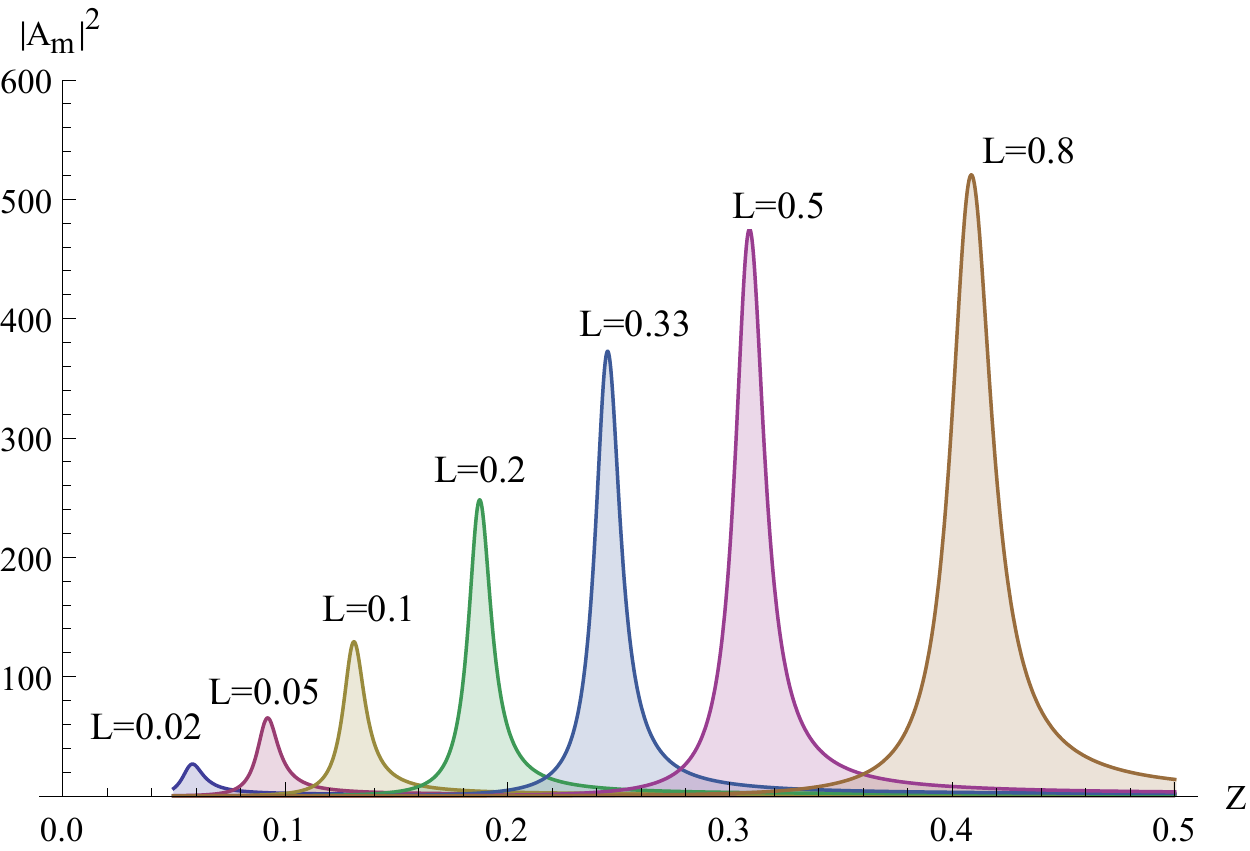}}
\vspace{-2mm}
	\caption{(Color online) The enhancement factor $|A_{\text m}|^{2}$ for a small silver particle versus a function of the dimensionless frequency $z$ at different $L$ similar to figure~\ref{fig-smp2} but with $\varepsilon_{\text m} = 6$.} \label{fig-smp3}
\end{figure}

As it is shown in figure~\ref{fig-smp3},  by changing $ \varepsilon_{\text m}$ and $L$, one can obtain even larger
$|A_{\text m}|^{2}$ which is around 500. This means that, at comparatively
large applied fields $E_{0}$ in the vicinity of the corresponding
plasma resonance, it is necessary to consider the nonlinear terms in
the dielectric function of equation~(\ref{eqn-3.3}).
Further, we will compare it with $|A|^{2}$ of metal covered
inclusions with different dielectric cores. This quantity is
calculated with the help of the enhancement factor of equation~(\ref{eqn-3.4})  with typical numerical
values of the dielectric functions of a composite.

\begin{figure}[!t]
	\centerline{\includegraphics[width=0.9\textwidth]{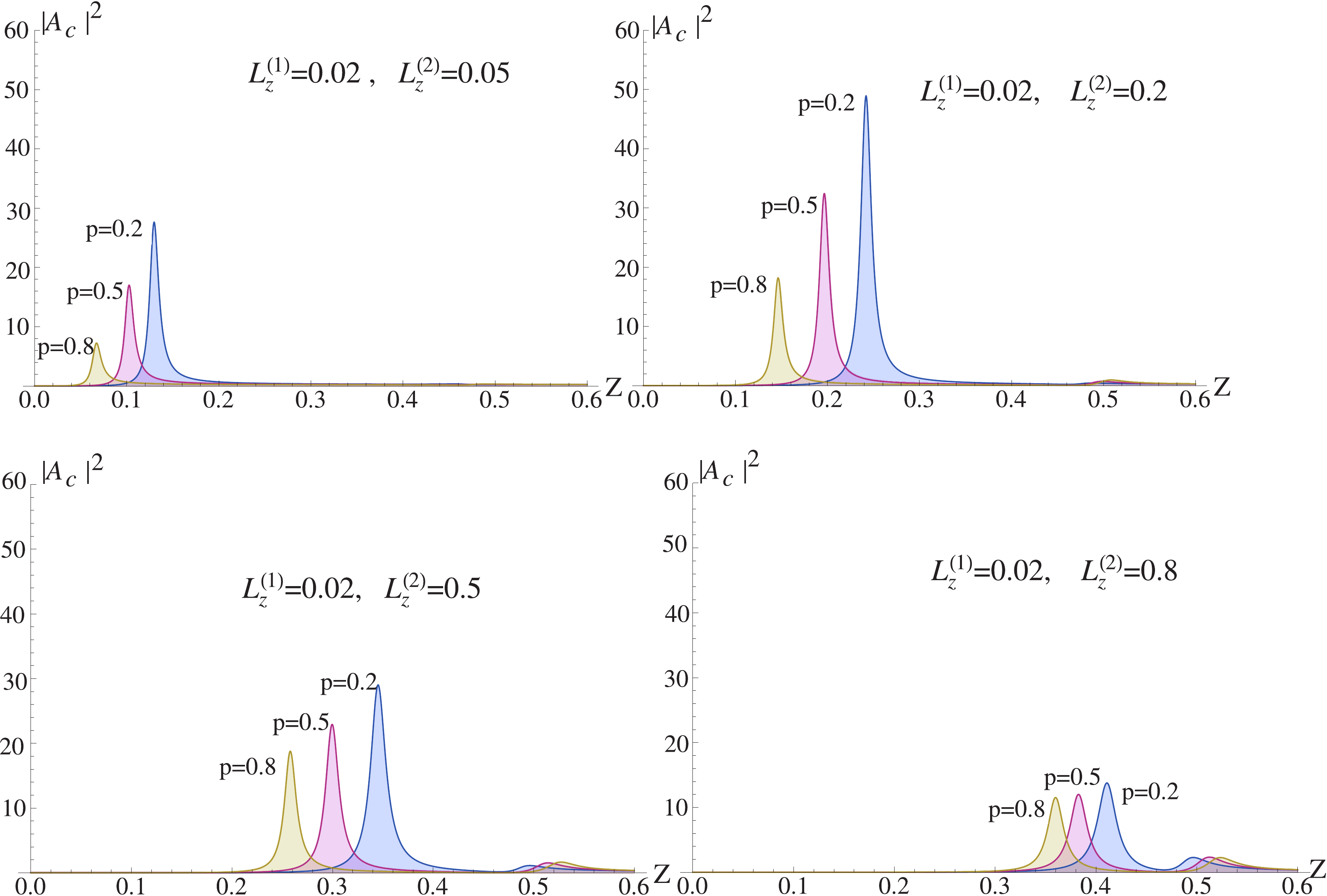}}
	\caption{(Color online) The enhancement factor $|A_{\text c}|^{2}$ for a silver
		ellipsoidal coated nanoparticle versus~$z$ for
		$L_{z}^{(1)}$(constant but minimum) $<$ $L_{z}^{(2)}$(variable), at $p = 0.2$, $0.5$, $0.8$. Here and further we use the following parameters: $\varepsilon_{\infty} = 4.5$,
		$\varepsilon_{1} = 6$, $ \varepsilon_{\text m} = 2.25$.} \label{fig-smp4a}
\end{figure}
\begin{figure}[!t]
	\centerline{\includegraphics[width=0.9\textwidth]{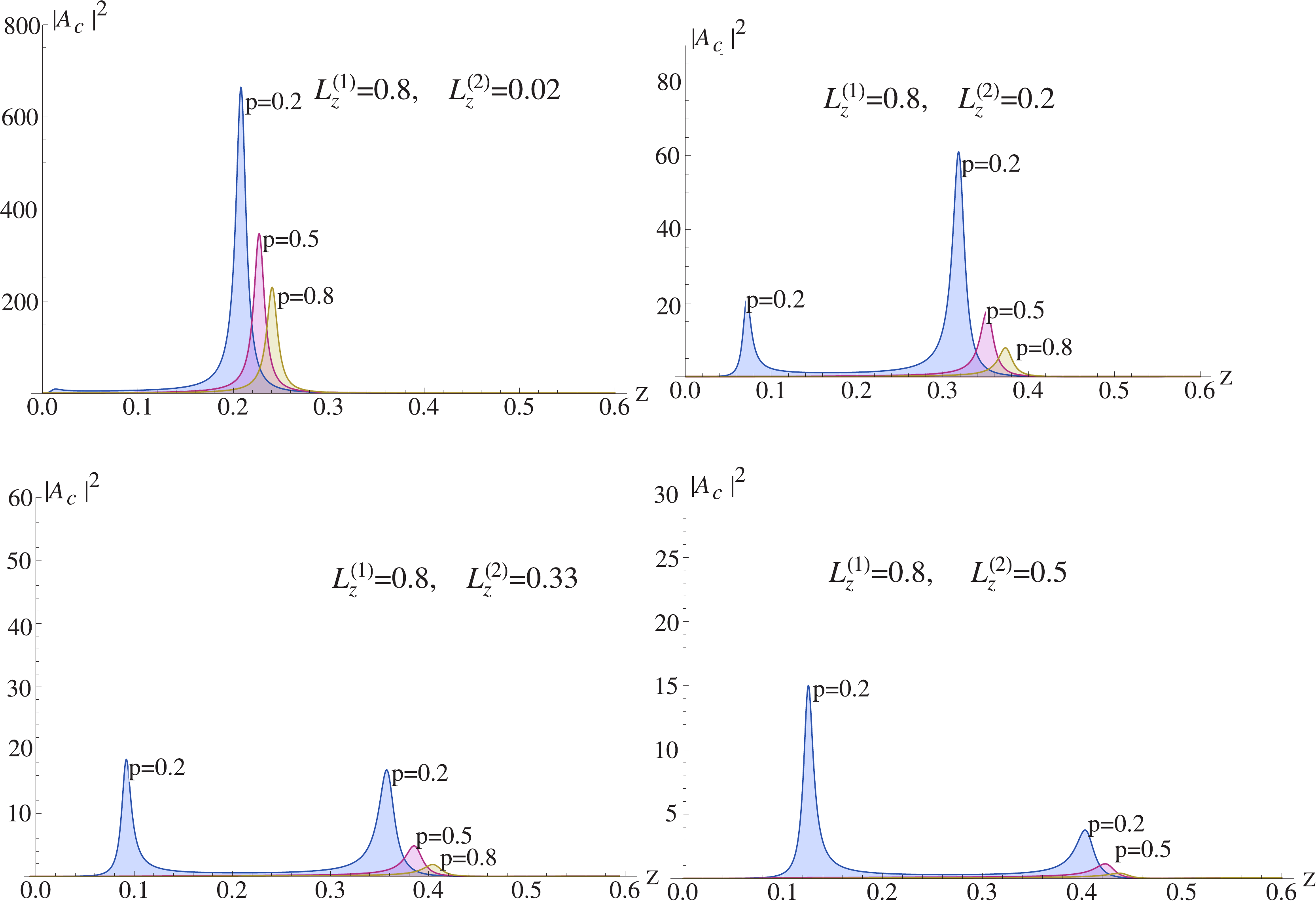}}
	\caption{(Color online) The enhancement factor $|A_{\text c}|^{2}$ for a silver
		ellipsoidal coated nanoparticle versus~$z$ for
		$L_{z}^{(1)}$(constant but maximum) $>$
		$L_{z}^{(2)}$(variable), at $p = 0.2$, $0.5$, $0.8$.} \label{fig-smp4b}
\end{figure}
\begin{figure}[!t]
	\centerline{\includegraphics[width=0.9\textwidth]{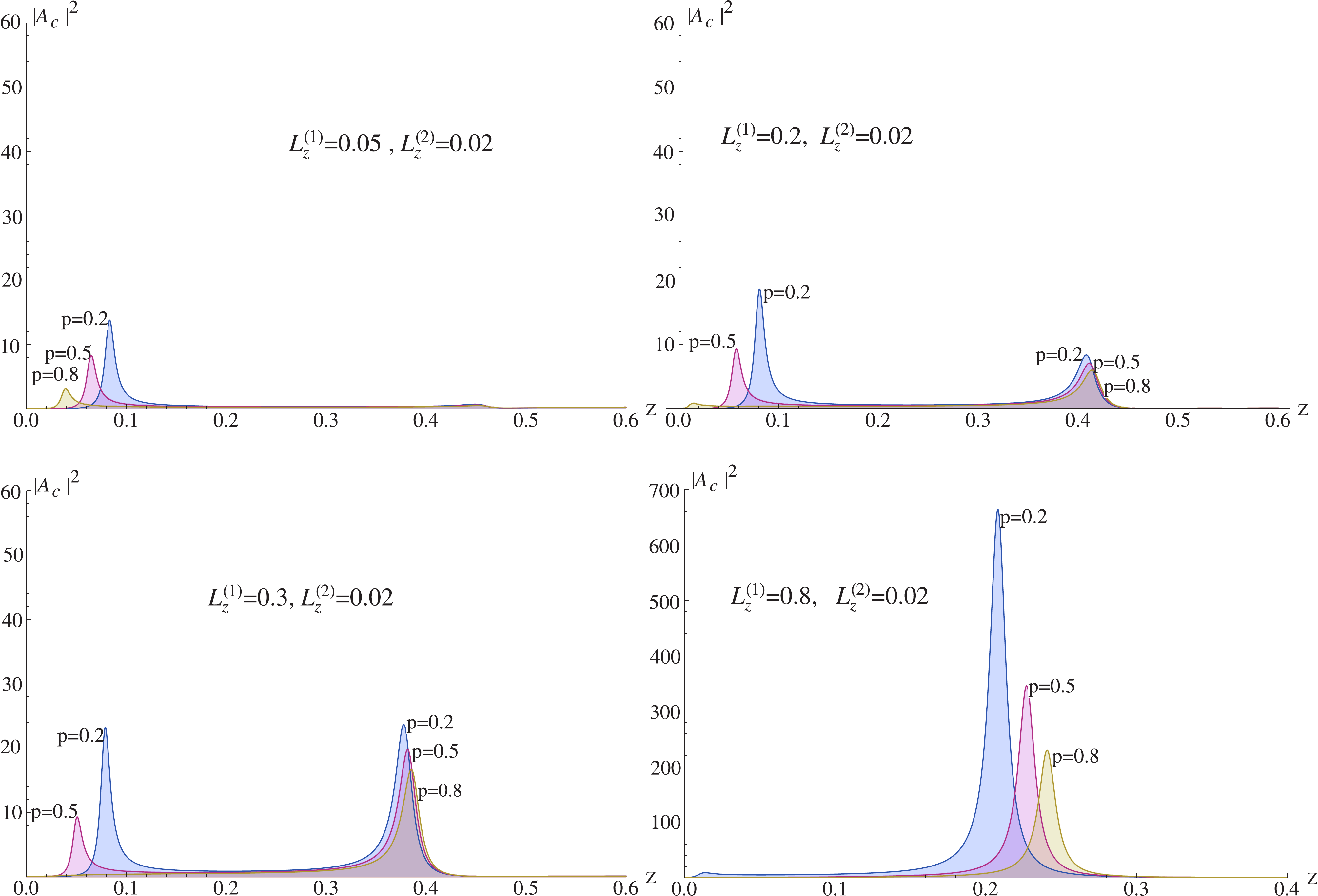}}
	\caption{(Color online) The enhancement factor $|A_{\text c}|^{2}$ for a silver
		ellipsoidal coated nanoparticle versus~$z$ for
		$L_{z}^{(2)}$(constant but minimum) $<$ $L_{z}^{(1)}$(variable).} \label{fig-smp5a}
\end{figure}
\begin{figure}[!t]
	\centerline{\includegraphics[width=0.9\textwidth]{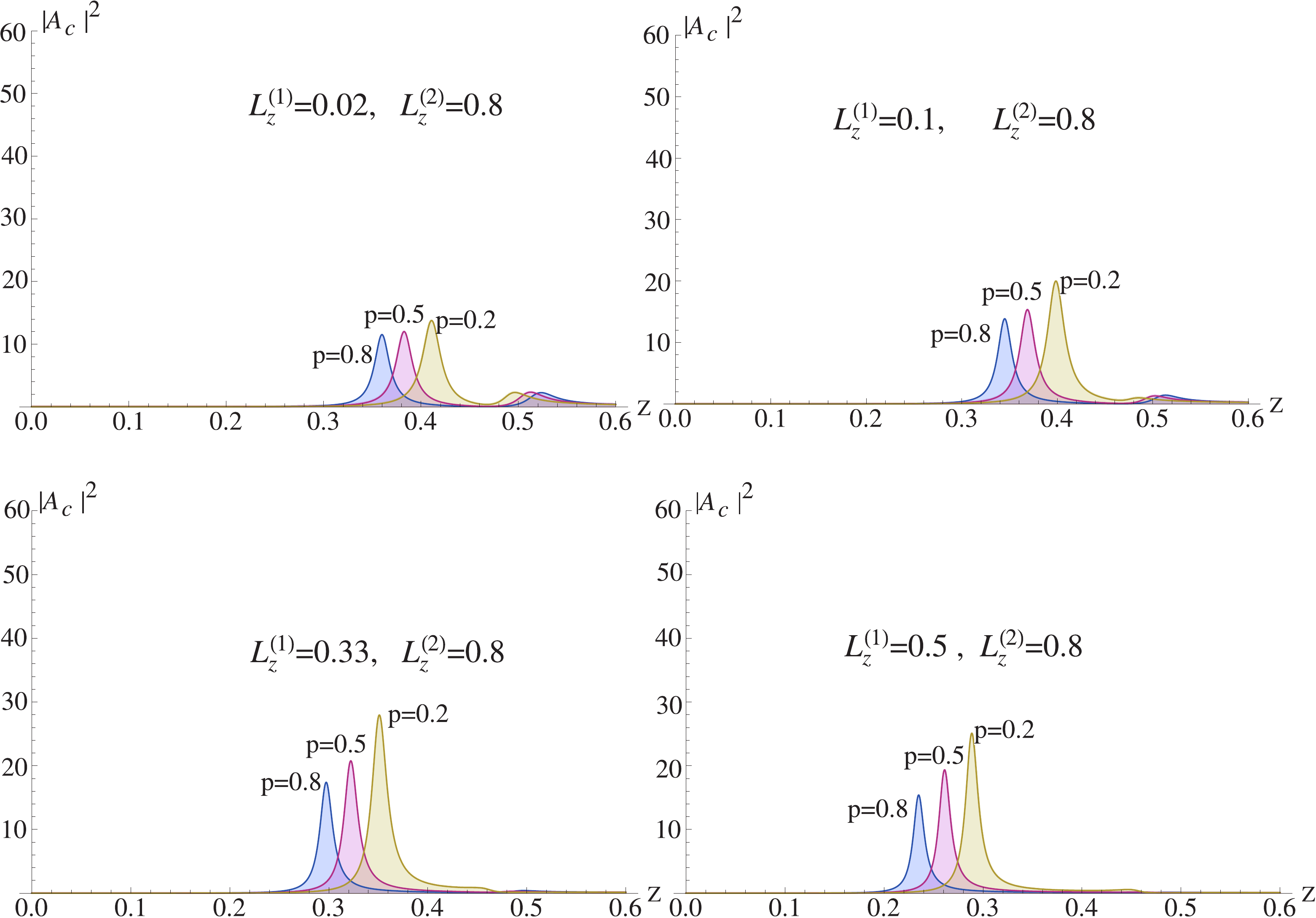}}
	\caption{(Color online) The enhancement factor $|A_{\text c}|^{2}$ for a silver
		ellipsoidal coated nanoparticle versus~$z$ for
		$L_{z}^{(2)}$(constant but maximum) $>$
		$L_{z}^{(1)}$(variable), at $p = 0.2$, $0.5$, $0.8$.} \label{fig-smp5b}
\end{figure}

Figures~\ref{fig-smp4a}--\ref{fig-smp5b} show the dependence of $|A|^{2}$ on the depolarization factor of the shell and the core at different metal fraction ($p$), respectively. Keeping the minimum value of the depolarization factor of the \emph{core} constant, it is observed that the enhancement factor $|A|
^{2}$ weakly depends on the depolarization factor of the \emph{shell}
decreasing with $L_{z}^{(2)}$, and the second maximum will appear as shown in figure~\ref{fig-smp4a}. At this instant, the extent of $|A|^{2}$ value for the small core (large $p$) becomes comparable with the large core (small $p$) of the inclusion.
When the maximum value of the \emph{core} depolarization factor is kept at
constant, as seen from figure~\ref{fig-smp4b}, the second maximum of the $|A|^{2}$ is dominant but
weakly depends on depolarization factor of the \emph{shell}, decreasing
with $L_{z}^{(2)}$ which in turn leaves the appearance on the
first maxima for the thin metal fraction of the inclusions.

Inspecting graphs in figure~\ref{fig-smp5a}, one can find that the second
maximum of the enhancement factor $|A|^{2}$ strongly depends on $L_{z}^{(1)}$ suppressing the result of the first maximum that
corresponds to a small \emph{core} (large $p$) which leads to its
disappearance at $L_{z}^{(1)}=0.8$ resulting in the $|A|^{2}$
value around 650 for a small metal fraction.

When $L_{z}^{(2)}=0.8$ is
kept constant, the dependance of the second maximum on the depolarization factor of the \emph{core} is not important,
while the first maximum increases with $L_{z}^{(1)}$ as shown in figure~\ref{fig-smp5b}.  Here, it should be emphasized that, unlike spherical case \cite{Sis12}, ellipsoidal inclusions having large dielectric core that exceeds the fraction  of metal (small $p$), the maximum value of
$|A|^{2}$ is
important depending on the suitable change of the depolarization
factor of the shell and the core.
This leads us to speak further about the composites of metal covered inclusions but not about the composites of dielectric inclusions having a metal core. Thus, we may presumably say that composites of metal covered inclusions behave in a manner different from the dielectric inclusions having a metal core.

\vspace{-1mm}
\section{Conclusion}

In this paper, we have shown that depolarization factors $L_{z}^{(1)}$ of the \emph{core} and $L_{z}^{(2)}$ of the \emph{shell} are the only factor that determines the magnitude of the enhancement factor. The enhancement factor of the local
field in a metal covered ellipsoidal inclusion with dielectric core
in a linear host matrix has two maxima at two different frequencies
that depends on the value of the core $L_{z}^{(1)}$ and the shell
$L_{z}^{(2)}$ depolarization factor. It may be noted that in sphere
the maxima are important in inclusions having large fraction of
metal ($p=0.9$) that exceeds the fraction of the dielectric \emph{core}, while in
our case, the maxima are important when the metal fraction ($p=0.2$) of ellipsoidal particle is very small.

\vspace{-1mm}
\section{Acknowledgements}

This work is dedicated to the late Professor V.N. Mal'nev
	who departed suddenly on January 22, 2015. We greatly acknowledge
	his invaluable contributions right from problem setting to almost
	its conclusion. Let his soul rest in peace.

\newpage
\ukrainianpart

\title{Пiдсилення локального електричного поля в текстурі кор-оболонка елiпсоїдних
 наночастинок метал/дiелектрик }
\author[A.A. Ismail, A.V. Gholap, Y.A. Abbo]{A.A. Ізмаіл\refaddr{label1}, А.В. Голап\refaddr{label1}, Й.А. Аббо\refaddr{label2}}
\addresses{
\addr{label1} Фізичний факультет,~університет Аддіс-Абеби, м. Аддіс-Абеба,~Ефіопія
\addr{label2} Фізичний факультет,~університет Уоллега,~P~O~Box~395,~Некемте,~Ефіопія}

\makeukrtitle

\begin{abstract}
В цій статті показано, що коефіцієнт підсилення локального електричного поля в еліпсоїдних частинках з металевим покриттям, вставлених в діелектричну матрицю, має два максимуми при різних частотах. Другий максимум для включень з металічним покриттям  з великим діелектричним кором (мала фракція металу $p$) є порівняно великим. Цей максимум сильно залежить від коефіцієнту деполяризації кору $L_{z}^{(1)}$, коли $L_z^{(2)}$ для оболонки залишається постiйним i меншим нiж $L_z^{(1)}$. Якщо частота зовнішнього випромінювання наближається до частоти поверхневих плазмонів металу, локальне
поле в частинці значно зростає.  Наголошується, що максимальне значення  коефіцієнта підсилення $|A|^{2}$ еліпсоїдального включення стає особливо важливим у випадку, коли діелектричний кор  перевищує металеву фракцію включення. Графічно представлені результати числових обчислень для типових малих срібних частинок.
\keywords коефіцієнт підсилення, еліпсоїдальне включення, коефіцієнт деполяризації, локальне поле, резонансна частота
\end{abstract}


\vspace{-1mm}
 \begin{thebibliography}{99}
 	\bibitem{Ref1} Prokes S.M., Glembocki O.J., Rendell R.W., Ancona M.G., Appl. Phys. Lett., 2007, \textbf{90}, 093105, \\ \bibdoi{10.1063/1.2709996}.
 	\bibitem{Ref2} Liu Y., Wang Sh., Park Y.-Sh., Yin X., Zhang X., Opt. Express, 2010, \textbf{18}, 25029, \bibdoi{10.1364/OE.18.025029}.
  	\bibitem{Ref4} Quan Q., Bulu I., Lon\v car M., Phys. Rev. A, 2009, \textbf{80}, 011810, \bibdoi{10.1103/PhysRevA.80.011810}.
 	\bibitem{Ref5} Russell K.J., Liu T.-L., Cui S., Hu E.L., Nat. Photonics, 2012, \textbf{6}, 459--462, \bibdoi{10.1038/nphoton.2012.112}.
 	\bibitem{Ref6} Spillane S., Kippenberg T., Vahala K., Nature, 2002, \textbf{415}, 621--623, \bibdoi{10.1038/415621a}.
 	\bibitem{Ref7} Van Thourhout D., Roels J., Nat. Photonics, 2010, \textbf{4}, 211--217, \bibdoi{10.1038/nphoton.2010.72}.
 	\bibitem{Ref8} Anker J.N., Hall W.P., Lyandres O., Shah N.C., Zhao J., Van Duyne R.P., Nat. Mater., 2008, \textbf{7}, 442--453, \bibdoi{10.1038/nmat2162}.
    \bibitem{Ref9} Juan M.L., Righini M., Quidant R., Nat. Photonics, 2011, \textbf{5}, 349--356, \bibdoi{10.1038/nphoton.2011.56}.
 	\bibitem{Nee89} Neeves A.E., Birnboim M.H., J. Opt. Soc. Am. B: Opt. Phys., 1989, \textbf{6}, 787, \bibdoi{10.1364/JOSAB.6.000787}.
 	\bibitem{Ref10}  Kalyaniwalla N., Haus J.W., Inguva R., Birnboim M.H., Phys. Rev. A, 1990, \textbf{42}, 5613,\\ \bibdoi{10.1103/PhysRevA.42.5613}.
 	\bibitem{Ref11} Haraguchi M., Okamoto T., Inoue T., Nakagaki M., Koizumi H., Yamaguchi K., 	Lai C., Fukui M., Kamano M., Fujii M., IEEE J. Sel. Top. Quantum Electron., 2008, \textbf{14}, No.~6, 1540, \bibdoi{10.1109/JSTQE.2008.917030}.
 	\bibitem{Bur11} Buryi O.A., Grechko L.G., Mal'nev V.N., Shewamare S., Ukr. J. Phys., 2011, \textbf{56}, 311.
    \bibitem{Sis12} Shewamare S., Mal'nev V.N., Physica B, 2012, \textbf{407}, 4837--4842, \bibdoi{10.1016/j.physb.2012.08.007}.
    \bibitem{url1} Wikipedia, Ellipsoidal coordinates --- {W}ikipedia{,} the free encyclopedia, 2016, [Online; accessed 10-Jan-2017], URL~\url{https://en.wikipedia.org/w/index.php?title=Ellipsoidal_coordinates&oldid=722351999}.
 	\bibitem{Boh83} Bohren C.F., Huffman D.R., Absorption and Scattering of Light by Small Particles, John Wiley, New York, 1983.
 	\bibitem{Abb16} Abbo Y.A., Mal'nev V.N., Ismail A.A., Condens. Matter Phys., 2016, \textbf{19}, No.~\textbf{3}, 33401,\\ \bibdoi{10.5488/CMP.19.33401}.

 \end{thebibliography}
 \end{document}